# Modular architecture facilitates noise-driven control of synchrony in neuronal networks

Hideaki Yamamoto[1,13*], F. Paul Spitzner[2,13], Taiki Takemuro[1,3], Victor Buendía[4,5,6], Carla Morante[7,8], Tomohiro Konno[9], Shigeo Sato[1], Ayumi Hirano-Iwata[1,3,10], Viola Priesemann[2,11,14], Miguel A. Muñoz[6,12,14], Johannes Zierenberg[2,14], Jordi Soriano[7,8,14*]

1. Research Institute of Electrical Communication (RIEC), Tohoku University, Sendai, Japan.
2. Max Planck Institute for Dynamics and Self-Organization, Göttingen, Germany.
3. Graduate School of Biomedical Engineering, Tohoku University, Sendai, Japan.
4. Max Planck Institute for Biological Cybernetics, Tübingen, Germany.
5. Department of Computer Science, University of Tübingen, Tübingen, Germany.
6. Departamento de Electromagnetismo y Física de la Materia, Universidad de Granada, Granada, Spain.
7. Departament de Física de la Matèria Condensada, Universitat de Barcelona, Barcelona, Spain.
8. Universitat de Barcelona Institute of Complex Systems (UBICS), Barcelona, Spain.
9. Graduate School of Pharmaceutical Sciences, Tohoku University, Sendai, Japan.
10. Advanced Institute for Materials Research (WPI-AIMR), Tohoku University, Sendai, Japan.
11. Institute for the Dynamics of Complex Systems, University of Göttingen, Göttingen, Germany.
12. Instituto Carlos I de Física Teórica y Computacional, Universidad de Granada, Granada, Spain.

13. These authors contributed equally to this work: Hideaki Yamamoto, F. Paul Spitzner

14. These authors jointly supervised this work: Viola Priesemann, Miguel A. Muñoz, Johannes Zierenberg, Jordi Soriano

*Email: hideaki.yamamoto.e3@tohoku.ac.jp; jordi.soriano@ub.edu

**ABSTRACT:**

Brain functions require both segregated processing of information in specialized circuits, as well as integration across circuits to perform high-level information processing. One possible way to implement these seemingly opposing demands is by flexibly switching between synchronous and less synchronous states. Understanding how complex synchronization patterns are controlled by the interaction of network architecture and external perturbations is thus a central challenge in neuroscience, but the mechanisms behind such interactions remain elusive. Here, we utilise precision neuroengineering to manipulate cultured neuronal networks and show that a modular architecture facilitates desynchronization upon asynchronous stimulation, making external noise a control parameter of synchrony. Using spiking neuron models, we then demonstrate that external noise can reduce the level of available synaptic resources, which make intermodular interactions more stochastic and thereby facilitates the breakdown of synchrony. Finally, the phenomenology of stochastic intermodular interactions is formulated into a mesoscopic model that incorporates a state-dependent gating mechanism for signal propagation. Taken together, our results demonstrate a network mechanism by which asynchronous inputs tune the inherent dynamical state in structured networks of excitable units.

**MAIN TEXT:**

The mammalian brain is in a state of perpetual ongoing activity characterized by high levels of irregularity in single-neuron response[1,2] and correlated fluctuations across brain regions[3–7]. Understanding the origin and functional significance of such neural activity has been challenging for both physics and neuroscience, and diverse competing hypotheses have been proposed to rationalize its nature. A compelling hypothesis in statistical physics is that the cortical networks operate nearby a critical point, e.g., at the edge of a non-equilibrium phase transition.[8–15] One possible transition the networks undergo is that between synchronous and asynchronous phases:[16,17] The synchronous phase could be argued to enable robust information propagation integrated across distance and time, while the asynchronous phase could enable segregated processing in local circuits with reduced redundancy. Flexible switching between these phases would enable networks to transiently exploit diverse functional advantages.

Transitions between synchronized and desynchronized states can be induced depending on the nature of external inputs[18,19] or by changes in connectivity strengths and network architecture[20–23]. Indeed, such transitions have been found in the mammalian brain. For instance, the thalamus projects asynchronous background inputs to the cortex,[24–26] which intuitively decrease the level of synchrony. Consistent with this, anaesthesia, which decreases the thalamocortical input,[27] enhances neural synchrony in the rat somatosensory cortex.[28] The deprivation of such inputs by anatomical lesions has also been shown to increase cortical synchrony and generate epileptic seizure-like activity in slice preparations.[29] Meanwhile, theoretical studies show that the response of a generic network to external perturbations strongly depends on the network architecture and the interaction strength.[30] Therefore, given that cortical networks are non-random[31] and exhibit strong modularity[5,32–34], it is reasonable to hypothesize that cortical dynamics rely on the asynchronous input they constantly receive from subcortical areas, such as the thalamus, in conjunction with the underlying network architecture. Despite accumulated evidence, the driving mechanisms that allow cortical networks to transiently regulate their level of synchronization remains elusive, both theoretically and experimentally.

Here we use *in vitro* cortical networks grown on engineered substrates[35] to address this question in a bottom-up approach. Three different types of networks with diverse degrees of modularity were grown, and their responses to optogenetic stimulation were assessed using fluorescence calcium imaging. The results show that modularity, together with asynchronous external input, enhance the dynamical repertoire by fostering local desynchronization in a GABA-dependent manner. The results are then compared with a spiking neural network model to show that a combination of stochastic intermodular interactions and decreased level of available synaptic resources rationalize the underlying mechanisms behind the experimental observations. Finally, we derive a mesoscopic model incorporating a state-dependent gating of intermodular interactions that minimally explains the input-driven control of synchrony. Taken together, our findings demonstrate a potential network mechanism by which asynchronous input serves as a control parameter that tunes the dynamical state inherent in structured neuronal networks.

**Disruption of synchrony by optogenetic stimulation**

We first assessed how external perturbations influence synchronized neural activity in biological neuronal networks grown *in vitro.* We designed modular micropatterns consisting of four small squares (200×200 $\mu m^2$) with connection lines that allowed a fraction of the neurites to interconnect the squares (Fig. 1a). The neural activity of the micropatterned networks was recorded by fluorescence calcium imaging using the calcium probe GCaMP6s (Fig. 1b, Supplementary Videos 1 and 2). Neurons were perturbed either by irradiating patterned light to individual neurons transfected with the photoactivatable cation channel ChrimsonR (optogenetic stimulation; Fig. 1c) or by increasing the extracellular potassium concentration $[K^+]_o$ (chemical stimulation). The former induces spiking activity in targeted neurons,[36] whereas the latter increases the spontaneous firing rate of neurons in the entire culture, effectively raising the overall neuronal excitability.[37]

In the non-stimulated state (Fig. 1d), the activity of the cultures was characterized by quasi-periodic episodes of network-wide bursting activity with some variability in amplitude due to the modular architecture.[35] External perturbations via optogenetic stimulation induced a qualitative change in network dynamics. When a spatiotemporally random illumination pattern was delivered to two out of the four modules (Fig. 1a), neurons in the network not only increased their activity but also exhibited a richer variety of collective activity patterns (Fig. 1e). This change in network dynamics was observed while stimulation was maintained and diminished when it was switched off (Fig. 1f). Representative snapshots of network behaviour before, during, and after stimulation are provided in Fig. 1g, illustrating the shift in collective activity from a synchronized to a desynchronized state upon stimulation. Chemical stimulation, in contrast, imposed a qualitatively different change in network dynamics (Fig. 1h,i; Supplementary Videos 3 and 4). Contrary to optogenetic stimulation, network-wide collective activity remained dominant, even in the perturbed state. These results indicate that a mere increase in excitability was insufficient and that an asynchronous stimulation was necessary to break synchrony and increase the dynamical repertoire of cortical networks.

Changes in collective activity during stimulation were quantified by measuring the distribution of event sizes, i.e., the fraction of neurons entrained in each collective activity episode (Fig. 1j). Optogenetic stimulation led to a significant decrease in event size due to the loss of synchrony. This change was accompanied by a broadened distribution of pairwise correlation coefficients, which manifested in an increasing trend of functional complexity (Fig. 1k), a signature of integration–segregation balance.[38] Chemical stimulation, however, preserved synchrony in network dynamics and showed the opposite trend in both the event size and functional complexity.

The overall effect induced by optogenetic stimulation was abolished in the presence of bicuculline (20 µM), a $GABA_A$ receptor antagonist that blocks inhibitory synapses, and the effect was thus GABA-dependent (Supplementary Fig. 1 and Videos 5,6). This indicates that GABAergic balancing of excitation and inhibition[39,40] is required for external input to alter network dynamics. This observation suggests that when a network is in an exceedingly excitable state, neurons are mostly depleted of neurotransmitters between collective activity events,[41,42] leading to a state that is insensitive to perturbations.

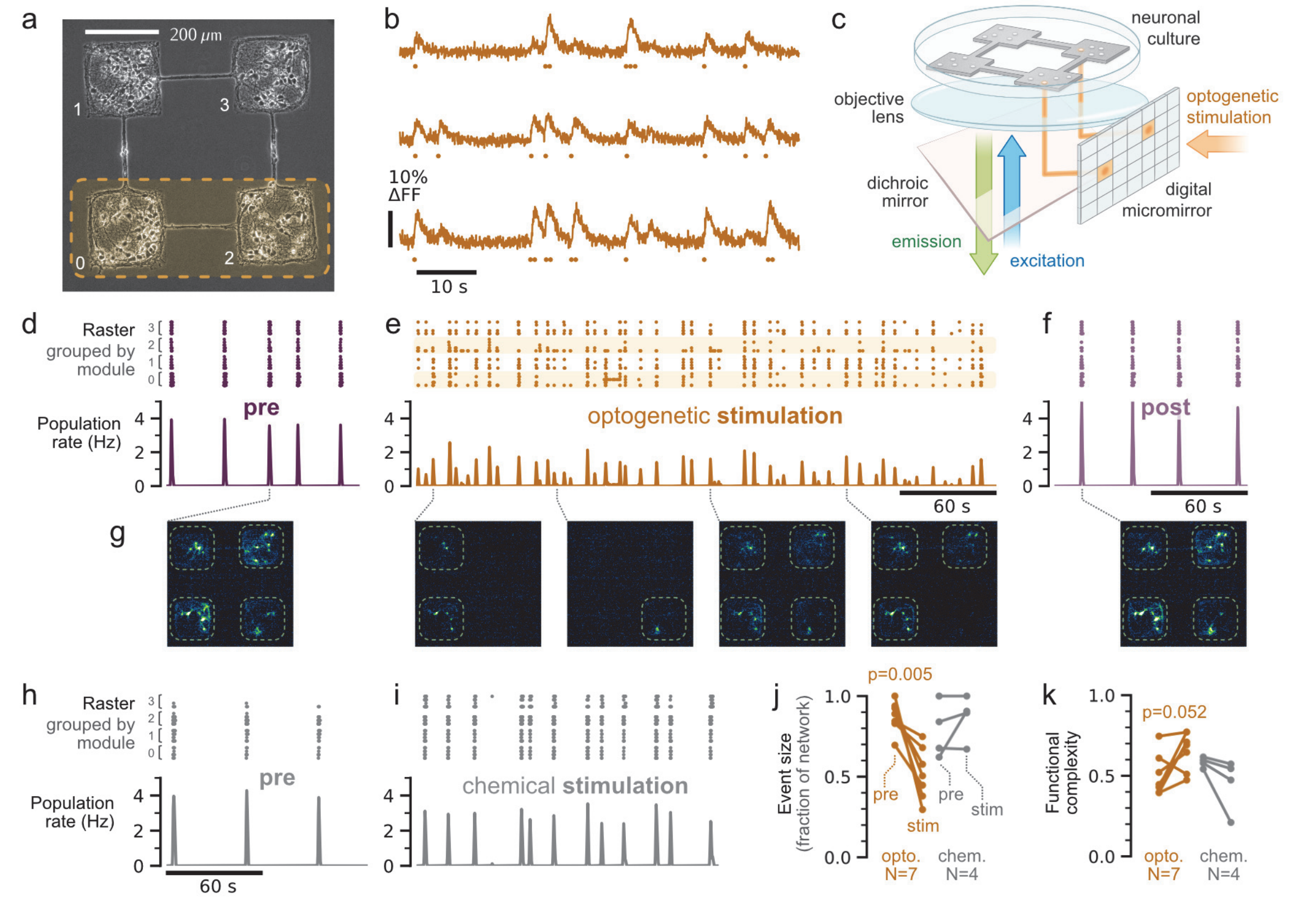

**Fig. 1. Optogenetic stimulation on modular neuronal cultures increases the variability in collective network**

**dynamics. (a)** Phase-contrast image of a representative single-bond modular network. Neurons appear as dark round objects with a white contour. Ten neurons were selected from the bottom module pair (orange box) and optogenetically targeted in a random manner. **(b)** Representative fluorescence traces and inferred spike events (dots) of three neurons along 1 min. **(c)** Sketch of the experimental setup. Neuronal cultures were transfected with ChrimsonR for optogenetic stimulation (orange arrow) and GCaMP6s for simultaneous activity monitoring (blue and green arrows). **(d)** Pre-stimulation raster plot (top panel) of network spontaneous activity, with neurons grouped according to their module, and the corresponding population activity (bottom panel). **(e)** Corresponding data upon optogenetic stimulation, wherein population activity markedly increases in variability. Targeted modules are marked as orange bands. **(f)** Spontaneous activity post-stimulation, with a return to strong network-wide bursting. **(g)** Representative snapshots of calcium imaging recordings for the above data. All modules activate synchronously without stimulation. Upon stimulation, activity events extend over individual neurons, multiple modules, or all modules. **(h–i)** Raster plot and population activity before and during chemical stimulation. Chemical stimulation increases the frequency of events but maintains the network-wide activity. **(j–k)** Effect of optogenetic and chemical stimulation on bursting event sizes and functional complexity (paired-sample *t*-test, two-sided). For chemical stimulation with $N = 4$, no test was performed.

**Impact of modular architecture**

Next, we assessed how the network architecture, specifically modular organization, interacts with external inputs to define network states. Hence, we prepared three types of networks with a constant number of neurons and different degrees of modularity (Fig. 2, top drawings; Supplementary Fig. 2). The modular micropattern used in the aforementioned experiments is hereafter referred to as the 'single-bond' (1-b) micropattern. Similarly, a 'triple-bond' (3-b) micropattern was designed by increasing the number of connection lines to three. Finally, the 'merged' micropattern was a single square of 400×400 $\mu m^2$. The modularity of the network, defined as the fraction of intermodular connections within a network greater than the expected fraction in a random network, decreased in the order of 1-b, 3-b, and merged.[35,43] A comparison of the distribution of event sizes and correlation coefficients revealed that cortical networks became less sensitive to perturbations when the modularity decreased. As shown in Fig. 2a, the decrease in median event size via optogenetic stimulation was 54% for the 1-b network, whereas the values were 21% and 25% for the 3-b and merged networks, respectively. A similar trend in structure dependence was also observed for the correlation coefficients (Fig. 2b), which decreased by 49%, 13%, and 19% for the 1-b, 3-b, and merged networks, respectively. These tendencies were also evident in the sample-averaged values (Supplementary Fig. 2).

To understand the mechanism of this structure-dependence, we analysed the shift in the correlation coefficient between neurons *i*-*j*, $r_{ij}$, during perturbation (Fig. 2c). For the 1-b network, the decrease in correlation largely stemmed from the neuron pairs that included at least one stimulated neuron, in which case $r_{ij}$ broadly scattered below the unit line in the pre-stim plane. Mean $r_{ij}$ strongly decreased when either one or both neurons in a pair were located within the stimulated modules, a feature that was not observed for neuron pairs in the modules that did not directly receive stimulation (Fig. 2d). Such clear spatial dependence was not observed when modularity was low (3-b, Fig. 2e) or absent (merged). Finally, the broadened distribution of correlation coefficients during stimulation increased the value of functional complexity, and the value was maximal in the 1-b network under stimulation (Fig. 2f). Summarizing, modularity fosters local and transient decorrelation from asynchronous stimulation, enabling cortical networks with sufficiently high modularity to support the coexistence of various degrees of synchronization.

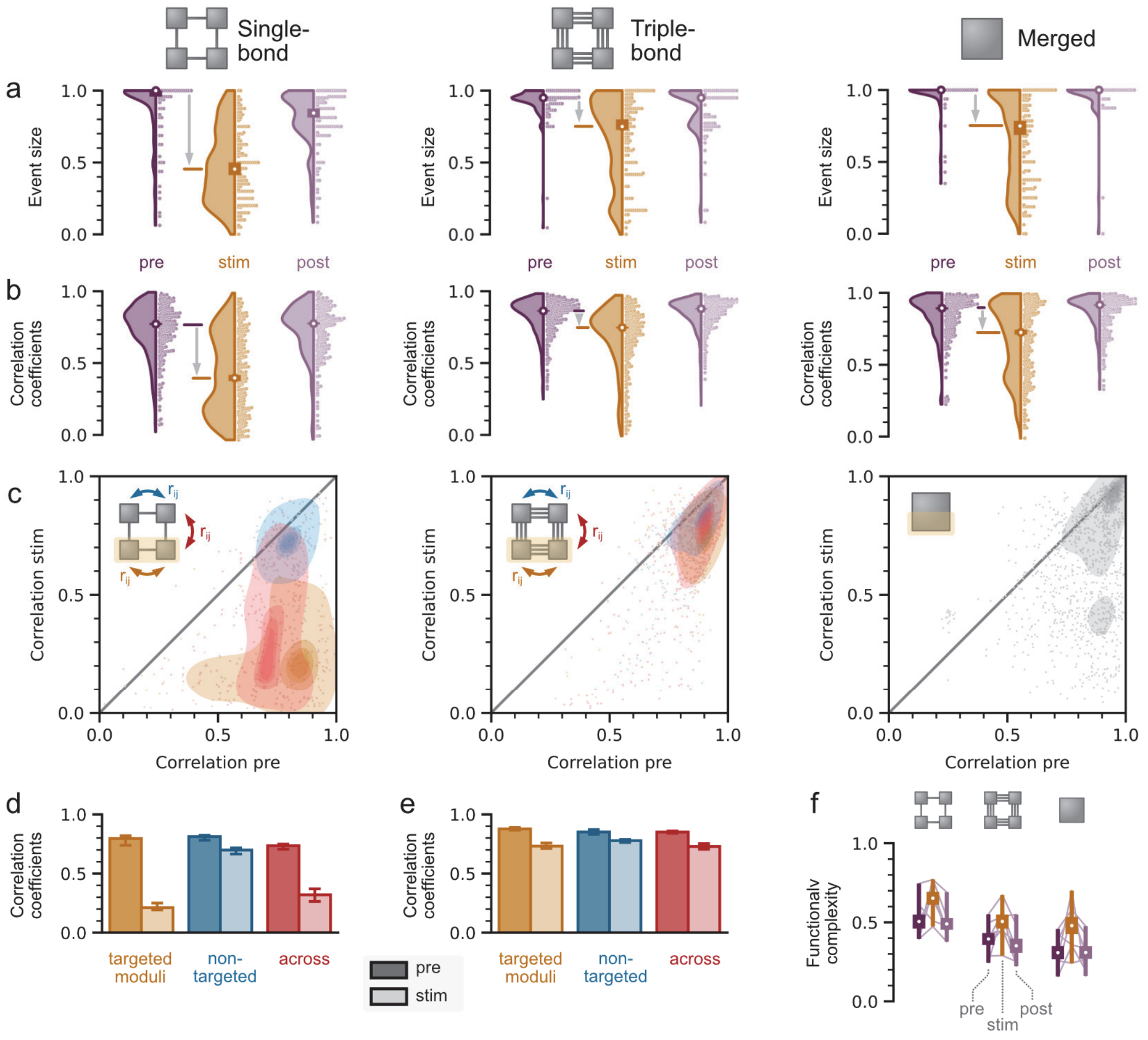

**Fig. 2**. **Disruption of network-wide collective activity upon optogenetic stimulation is facilitated by modular architecture. (a)** From left to right, event size distribution before, during, and after optogenetic stimulation for the 1-b, 3-b, and merged networks. The 1-b network exhibits a drop of ~50% in average event size, contrasting with the ~20% drop for 3-b and merged. Violin plots (left side) represent smooth kernel estimates of the events observed across all networks, while individual observations are shown in swarms (right side). Error bars (centre) are obtained via bootstrapping. White dots indicate the median of the 500 bootstrap estimates and bars represent the 95 percentiles. **(b)** Corresponding distribution of pairwise Pearson correlation coefficients between neurons. A substantial drop is only observed for 1-b. Data are presented as in panel (a). **(c)** Change of correlation coefficients $r_{ij}$ between the pre-stimulated and stimulated conditions. For 1-b and 3-b modular cultures, neuron pairs are grouped according to the modules in which they are located. Both neurons may either reside in modules that are targeted by stimulation

(yellow), both reside in non-targeted modules (blue), or the pair spans across a targeted and non-targeted module (red). Decorrelation is more pronounced when one or both neurons are in modules targeted by stimulation. Coloured areas are fitted probability density estimates for each data group. The diagonal black line is the reference condition in which no changes occur. **(d–e)** Median correlation coefficients for 1-b and 3-b modular networks. Error bars represent 95 percentiles. **(f)** Functional complexity values for the three topologies. White dot: mean of 500 bootstrap samples, thick bars: SEM, thin bars: extrema, thin lines: individual networks. 1-b: $p = 0.052$ (pre-stim), 0.004 (stim-post), 0.539 (pre-post); 3-b: $p = 0.038$ (pre-stim), 0.017 (stim-post), 0.067 (pre-post); merged: $p = 0.047$ (pre-stim), 0.056 (stim-post), 0.999 (pre-post) (paired-sample $t$-test, two-sided).

## Microscopic spiking neural network model

To rationalize the underlying mechanisms behind the enhanced sensitivity to external perturbations in modular networks, we next constructed a spiking neural network (SNN) model based on leaky integrate-and-fire neurons (see Methods and Supplementary Material Section S2 for details). The networks were generated based on the metric-construction approach described previously[44], and modularity was tuned by specifying the number of axons that crossed from one module to another (Fig. 3a). With intermodular connectivity $k = 5$, we obtained dynamic behaviours comparable to experiments in 1-b networks (Fig. 3b). The model accurately recapitulated the experimental observations: In the pre-stimulated state (Fig. 3b, left), spontaneous activity was driven by Poisson noise, representing the spontaneous release of neurotransmitters in biological presynaptic terminals,[41,44] which led to activity patterns that combined sporadic activity with network-wide bursting. Stimulation, introduced in the two lower modules as additional noise that effectively mimicked inward current pulses from optogenetic inputs (Fig. 3b, right), led to a clear breakdown of synchrony among modules. As shown in Fig. 3c, the median event size decreased from 0.8 to 0.1, and the median correlation (Fig. 3d) decreased from 0.75 to 0.3. Corroborating the experiments, a greater decrease in the correlation coefficient was observed for cell pairs wherein one of the cells belonged to the targeted modules (Fig. 3e–f), whereas the change was less pronounced for cell pairs in non-targeted modules.

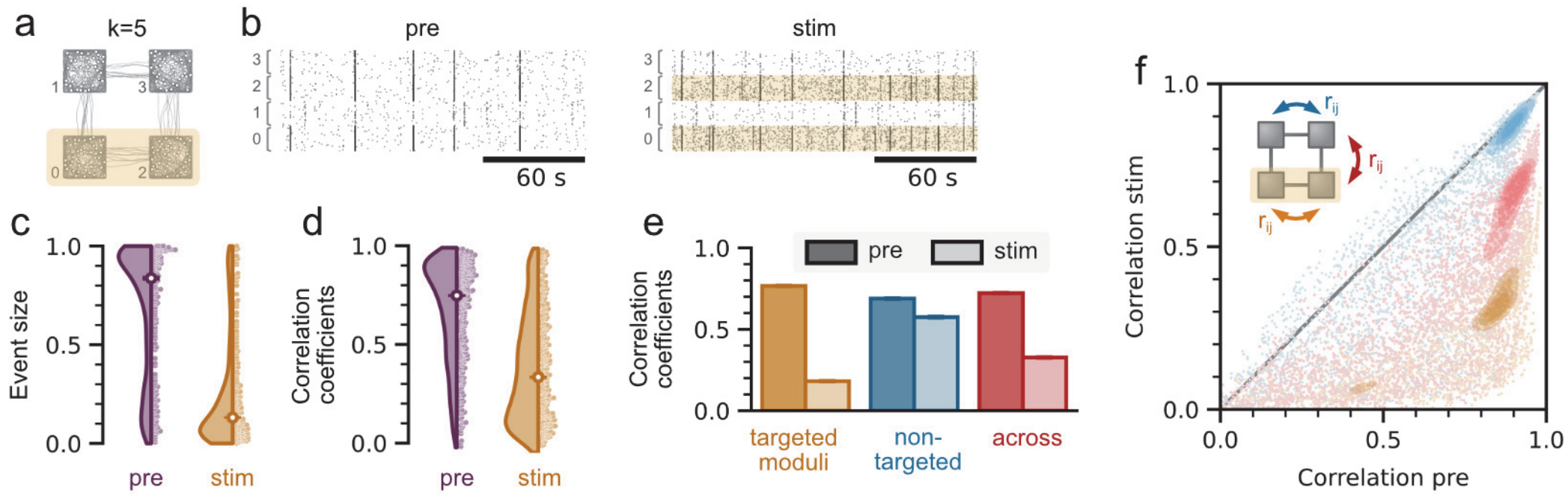

**Fig. 3.** ***In silico*** **simulations of modular networks replicate the decrease in burst event size and correlation coefficients**. **(a)** Sketch of a simulated modular network with $k = 5$ output connections from a module to its neighbours. **(b)** Representative raster plots in the pre-stimulated (left) and stimulated (right) regimes, illustrating the loss of network synchronous bursting for the latter. Orange boxes highlight stimulated modules. **(c)–(d)** Distributions of event size and pairwise correlation coefficients before and during stimulation. Both distributions exhibit a substantial drop towards smaller values upon stimulation. **(e)** Bar plots of pairwise correlation coefficients, comparing the behaviour of neurons in the same stimulation-targeted modules (yellow), non-targeted modules (blue), and pairs spanning targeted and non-targeted modules (red). **(f)** Corresponding plot of pairwise correlation coefficients ($r_{ij}$) between the pre-stimulated and stimulated regimes. The strongest decorrelation occurs in neuronal pairs wherein at least one of the neurons belongs to a targeted module. All data were obtained from identical analysis as in Fig. 2 and are presented in the same way.

## Network mechanisms: The importance of synaptic resources

Above we found that modules targeted by increased input decorrelate more than non-targeted modules. Thus, we next generalize the experimental paradigm by systematically modifying the rate of synaptic noise delivered to all neurons in all modules. Two representative simulations at low and high noise rates are illustrated in Fig. 4a,b. The low-noise case (Fig. 4a, Supplementary Video 7) mimicked the pre-stimulated state in which network-wide bursting and sporadic activity coexisted, while the high-noise case (Fig. 4b, Supplementary Video 8) resembled the stimulated condition with reduced synchronous bursting. Note the characteristic behaviour of the average synaptic resources $R \in [0,1]$ per module (Fig. 4a,b, bottom rows), which abruptly discharge during bursting events and gradually recharge between events. At a low noise rate, $R$ had a mean value of ~0.8, but dropped to ~0.3 concurrent with the emission of network-wide bursting events. In contrast, when neurons were subjected to high-rate noise, $R$ only recovered to ~0.6, effectively reducing the mean synaptic efficacy, and network-wide bursts lowered $R$ only down to ~0.4.

Increased noise rate reduced the capacity of the network to exhibit collective network-wide bursting events (Fig. 4c), with synchronous activity gradually shrinking from network to module scales. Concurrently, we observed an increase in the propagation delay between modules (insets of Fig. 4ab; Supplementary Fig. 4), which indicates that noise not only disrupts the occurrence of network-wide bursting but also its spatiotemporal structure.

For a detailed investigation of network modularity and perturbation altering network-wide synchronous events, we explored topologies with different degrees of modularity, i.e., modular networks with the number of axons crossing the modules $k = 1$, 5, 10, and a non-modular ('merged') network (Fig. 4d), with modularity $Q = 0.70$, 0.53, 0.32, and 0, respectively. The simulation results show that desynchronization caused by stimulation tightly interplays with the modularity of the underlying topology (Fig. 4e–g, Supplementary Fig. 4). As a general trend, an increased noise rate decreased the mean event size (Fig. 4e) and pairwise correlation (Fig. 4f). Depending on the noise level, the network activity underwent a change from highly correlated state, over a flat distribution with different levels of correlation, to an uncorrelated state (cf. Fig. 3d). At extremely high noise rates (>100 Hz), the networks were purely noise-driven, expressing no collective activity events, and pairwise correlations approached zero. The impact of network architecture was most apparent in the noise-dependence of functional complexity (Fig. 4g), whose peak shifted towards a lower noise rate as the network modularity increased. This was caused by the inherent suppression of network-wide bursts owing to the modular architecture. The non-modular merged network showed a high functional complexity only for high noise levels, an artificial situation wherein dynamics is purely noise-driven.

Importantly, we find that the combination of modularity and asynchronous stimulation strongly affected the charge–discharge dynamics of the synaptic resources $R$. To illustrate this, it is useful to consider trajectories of network dynamics in the resource–rate plane (Fig. 4h), where large cycles are associated with severe discharge episodes driven by synchronous events. Independent of the network architecture, increased noise decreased the overall size of the cycles. However, modularity affected the variability of cycle sizes between different modules. Close to disconnecting modules ($k = 1$), cycles reflected single-module properties and differed in size. For intermediate connections ($k = 5$), cycles tended to be smaller due to interactions between modules but retained module-dependent sizes. In this case, the effect of noise was strongest, causing very small charge-discharge cycles signalling the breakdown of synchrony. When intermodular connection was further increased ($k = 10$), cycles start to become indistinguishable between modules, similar to the merged case.

In summary, modular architecture is essential for a noise-induced breakdown of synchrony: The heterogeneous degree distribution, with more connections within modules than between modules, renders intermodular activity propagation more sensitive to the noise-induced reduction of average resources than intramodular activations.

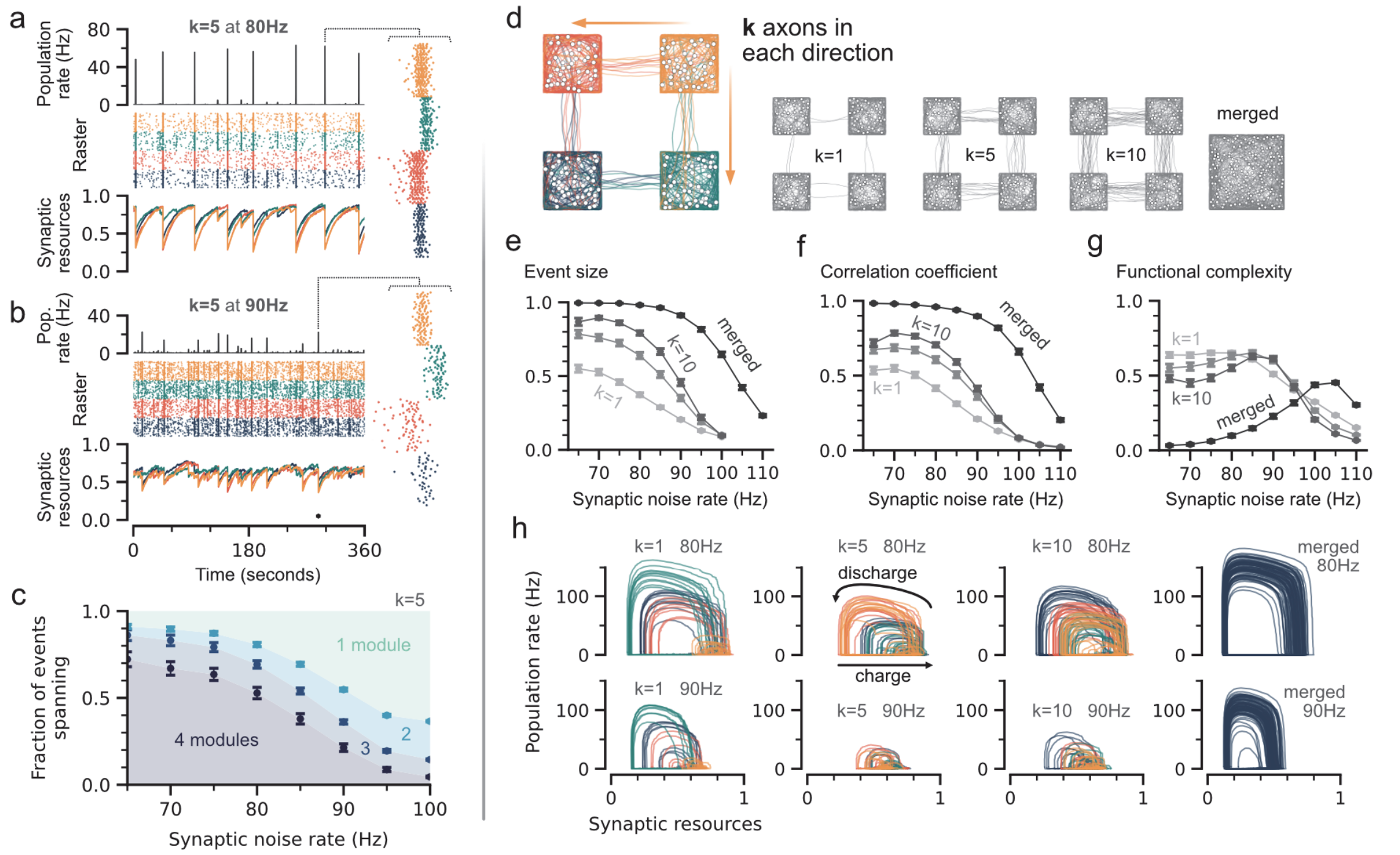

**Fig. 4. Disruption of network-wide synchronization upon enhanced synaptic noise *in silico* is mediated by the depletion of synaptic resources. (a)** Population rate, raster plot, and available synaptic resources for a modular network with mild synaptic noise of 80 Hz that represents the reference scenario without stimulation. Activity is dominated by synchronous bursting events, during which the synaptic resources are abruptly released and slowly recharge. A detail of the temporal structure of a bursting event is shown on the right. **(b)** Stimulated scenario, wherein the synaptic noise to all neurons is increased to 90 Hz. Network-wide events are less frequent and overall activity is more irregular. Synaptic resources fluctuate around an intermediate level, which hinders episodes of abrupt and severe discharge. **(c)** Fraction of observed bursting events that encompass 1, 2, 3, or 4 modules as a function of the noise strength. Network-wide events are dominant for low noise levels, while single-module events become dominant for high noise. **(d)** Sketches of *in silico* neuronal networks with $k = 1$, 5, and 10. The 'merged' topology has all four modules combined as a single network. **(e)** Network synchronization, quantified by the bursting event size. The stronger the coupling among modules, the higher the required noise to decrease event size. **(f)** Dependence of pairwise correlation on $k$ and noise strength. Activity is maximally correlated at different levels of synaptic noise. **(g)** Corresponding functional complexity analysis. In modular networks, functional complexity exhibits a maximum for intermediate noise levels. **(h)** Characteristic charge–discharge cycles of synaptic resources for the modular and non-modular networks. Top and bottom rows correspond to low- and high-noise cases, respectively. In general, cycle excursions were smallest for moderate modularity and increased noise. Representative movies of the simulations are provided in the Supplementary Material.

## Mesoscopic description

Finally, to extrapolate the microscopic behaviour of individual neurons to the macroscopic dynamics in the brain, we built a mesoscopic module-level model that captures the key empirical findings. Here, the dynamics of each module were described by a rate model with resource depletion,[12] which is defined by two coupled differential equations that represent the evolution of firing rate and synaptic resources, respectively. The latter accounts for the effect of short-term depression (see Methods and Supplementary Section 3 for details).

As a first exploration of the model, we considered the case in which each module received input through a non-linear activation function that depended on the rate and resources of the connected modules and an external input $h$ that captured the main (average) effect of stimulation. However, in such a case, increasing $h$ decreased the size of the resource–rate cycles of each module but did not affect synchronization (Supplementary Fig. 8), and dynamics desynchronized only when almost disconnecting the four modules by reducing their coupling strength.

Thus, we introduced non-deterministic intermodular interactions as 'gates' that stochastically disconnect when synaptic resources are depleted and reconnect after a characteristic time (Fig. 5a,b). These gates capture the essence of the microscopic dynamics, wherein intermodular coupling only operates when the synapses projecting from one module to another are not fully depleted. Numerical simulations revealed that this minimal modification was sufficient to recapitulate the noise-dependent breakdown of synchrony observed in the experiments and SNN model (Fig. 5c–f).

This simple mesoscopic approach allows us to investigate the interaction of stochastic intermodular interaction with input-dependent resource–rate cycles. The two components together form the fundamental mechanisms behind the noise-induced breakdown of synchrony in modular neuronal networks (see Supplementary Section 3). The mesoscopic model is also scalable and paves the way for computationally modelling the dynamics of large-scale networks using, e.g., human cortical connectomics data, with biologically-validated node models.

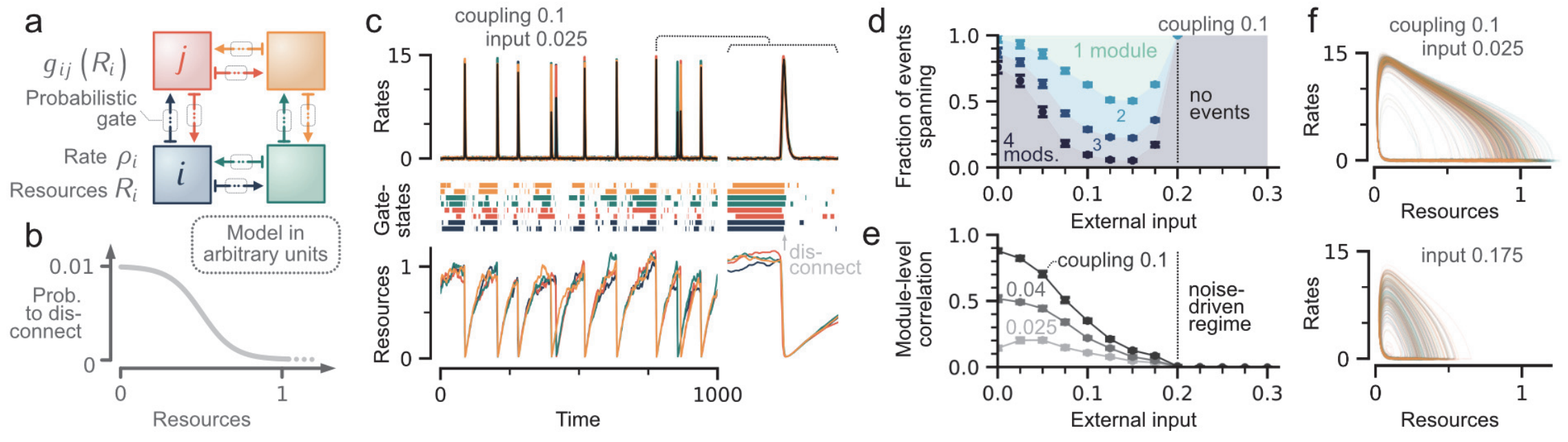

**Fig. 5. Mesoscopic model. (a)** Schematic representation of the model. Each module (coloured box) is treated as an effective unit described by rate and resource variables interacting with neighbouring units through probabilistic gates. **(b)** The probability of the gates to disconnect the modules is maximum when resources are depleted. **(c)** Firing rates of the 4-module network for a strong coupling and a weak external input, together with the states of the gates (coloured bars indicate open gates) and resources. Mesoscopic dynamics well reproduce the strong synchronous behaviour and abrupt depletion of resources observed in the microscopic SNN model. The right panels are a zoom-in. **(d)** An increase in external input at the same coupling strength shifts network behaviour toward a more asynchronous regime, until reaching a saturated regime where events are no longer detectable. **(e)** Module-level correlation as a function of the external input for different coupling strengths. Weak (0.025) and moderate (0.04) coupling substantially reduce the number of modules participating in events, leading to lower correlation values. **(f)** Rate–

resource cycles for strong intermodular coupling and weak (top) and strong (bottom) external input, with cycles substantially shrinking for the latter.

### Biological implications

Taking advantage of *in vitro* experiments using cultured neuronal networks as a model biological system,[35,44–47] we showed that the interaction of network architecture and asynchronous input plays a pivotal role in shaping the dynamical state of neuronal networks. Computational models revealed the underlying network mechanisms behind the observed phenomena. Our experiments also revealed that GABAergic inhibition is necessary for the neuronal network to modulate its level of synchrony, because when inhibition was blocked, synchronous epileptic-like activity prevailed (Supplementary Fig. 1). The same trend was also observed in the computational model of spiking neurons (Supplementary Fig. 5). The fundamental role of inhibition in shaping asynchronous states has been explored both experimentally and theoretically, revealing that networks of purely excitatory neurons are not able to generate stable asynchronous states.[17] This critical role of inhibition in stabilizing system dynamics highlights the evolutionary significance of preserving the balance of electrical–chemical signal transduction in the nervous system.

It has been conjectured that the function of neuronal networks requires the segregated processing of diverse inputs in specialized circuits as well as the integration of all of them to generate high-level information processing and response.[48] This demands a flexible balance of segregation and integration, the loss of which may induce dysfunction.[49] In dynamical terms, such an optimal balance is necessarily associated with high diversity and variability of underlying synchronization patterns of neural activity to be sustained.[50] Therefore, understanding how network-structural features and dynamical aspects collectively shape complex synchronization patterns is crucial for advances in the field. Our findings are also relevant to understand other networked systems that possess modular architectures and are subjected to noise, such as genes, epidemics, and social networks.

**METHODS:**

**Micropatterned substrate.** Microcontact printing was used to pattern protein ink onto glass coverslips. First, glass coverslips (C018001, Matsunami Glass Ind.) were cleaned by sonication in 100% ethanol, rinsed in Milli-Q water, and treated with air plasma for 60 s (PM-100, Yamato). The cleaned coverslips were then treated with a 0.2% solution of poly(2-methacryloyloxyethyl phosphorylcholine-co-3-methacryloxypropyl triethoxysilane)[51] in ethanol for 10 s, dried in an ethanol environment for 20 min, baked in an oven at 70 °C for 4 h, and dried under vacuum overnight. The coverslips were then sterilized by immersion in ethanol, rinsed in Milli-Q water, and dried. Protein ink [ECM gel (E1270, Sigma-Aldrich; 1:100 dilution) + poly-D-lysine (50 μg $ml^{-1}$; P0899, Sigma-Aldrich)] was patterned using a PDMS stamp. The fabrication of the polydimethylsiloxane (PDMS) stamp has been detailed previously[35]. Four pieces of thin PDMS films (approximately 2 mm × 2 mm and 0.5 mm thickness) were then attached to the periphery of the coverslip, which served as spacers. Finally, the coverslips were dried overnight in a fume hood and immersed in neuronal plating medium [minimum essential medium (MEM; 11095-080, Gibco) + 5% foetal bovine serum + 0.6% D-glucose].

**Cell culture.** The culture protocol of primary rat cortical neurons has been described previously[35,52]. Briefly, primary neurons were obtained from the cortices of embryonic day 18 pups, plated on a microfabricated coverslip at a density of 360–480 cells $mm^{-2}$, and co-cultured with astrocyte feeder cells in N2 medium containing MEM + N2 supplement + ovalbumin (0.5 mg $ml^{-1}$) + 10 mM HEPES]. Half of the medium was changed at DIV 4 and DIV 8 with a conditioned Neurobasal medium containing: Neurobasal (21103-049, Gibco) + 2% B-27 supplement (17504-044, Gibco) + 1% GlutaMAX-I (35050-061, Gibco). In some experiments, neurons were cultured in the Neuron Culture Medium (FujiFilm Wako Pure Chemical Corp. 148-09671), a glia-conditioned medium. The astrocyte feeder layer was not used when culturing the neurons in the latter medium.

During cultivation, neurons were transfected with the fluorescent calcium probe GCaMP6s (Addgene viral prep #100843-AAV9) and a red-shifted channelrhodopsin ChrimsonR (Addgene viral prep #59171-AAV9) using adeno-associated virus vectors. The as-received viral preparations were aliquoted and added at concentrations of 1 μL $mL^{-1}$ (GCaMP6s) and 0.7 μL $mL^{-1}$ (ChrimsonR) at DIV 4. The AAVs were diluted during medium exchange but remained in the growth medium until the end of the culture. All procedures were approved by the Tohoku University Center for Laboratory Animal Research, Tohoku University (approval number: 2020AmA-001) and Tohoku University Center for Gene Research (2019AmLMO-001).

**Calcium imaging.** At DIV 10–11, the coverslips with micropatterned neurons were rinsed in HEPES-buffered saline (HBS) containing 128 mM NaCl, 4 mM KCl, 1 mM $CaCl_2$, 1 mM $MgCl_2$, 10 mM D-glucose, 10 mM HEPES, and 45 mM sucrose, and transferred to a glass-bottom dish (3960-035, Iwaki) filled with HBS.[35,52] GCaMP6s fluorescence was imaged using an inverted microscope (Olympus IX83) equipped with a 20× objective lens (numerical aperture, 0.75), white-light light-emitting diode (LED) (Sutter Lambda HPX), sCMOS camera (Andor Zyla 4.2P), and stage-top incubator (Tokai Hit). All recordings were performed at 37 °C. Two networks were selected from a coverslip for the recording. A recording session of a network consisted of three phases: Phase 1 was a spontaneous activity recording, Phase 2 a recording with optogenetic stimulation (see below), and Phase 3 a spontaneous activity recording. Each phase lasted for 10 min, and time-lapse images were taken at 20 frames $s^{-1}$ using Solis software (Andor).

**Optogenetic stimulation.** Patterned light illumination for activating ChrimsonR was delivered using a digital mirror device (DMD) (Mightex Polygon400G) coupled to a high-power LED (Thorlabs Solis 623C; nominal wavelength, 623 nm) via a liquid light guide. The DMD was mounted on the inverted microscope, and patterned light was reflected onto the sample stage using a short-pass dichroic mirror with an edge frequency of 556 nm (Semrock FF556-SDi01). The spatiotemporal pattern of light illumination was designed in custom MATLAB script and programmed to the DMD using PolyScan2 software (Mightex). In the MATLAB script, somas of 10 neurons expressing ChrimsonR were randomly selected from the lower half of the cultured neuronal network. Subsequently, a circular illumination area centred around the soma (diameter, 25 µm) was generated randomly with a probability of 40% for each position. Finally, 750 black-and-white bitmap files with the illumination pattern were generated and imported into PolyScan2. The duration of each frame was set to 400 ms, which was sufficiently long to initiate one or more spiking activities in the illuminated neuron. Identical spatiotemporal patterns were repeated in the first and second halves of the 10 min session.

**Spike detection.** To extract the neural activity, ROIs were manually set around the neuronal somas using the CellMagicWand plugin in ImageJ2, and the mean intensity within the ROIs was extracted for each time step. ROIs with no activity were not used, and an equal number of neurons were selected from each of the four modules. Spikes were detected from calcium fluorescence traces using the MLSpike algorithm.[53] The first 1 min of each 10 min recording was removed to eliminate artifacts originating from the session onset. The algorithm occasionally detected pulse signals originating in the residual stimulation light as spikes, which were manually inspected and removed based on their shape and duration.

**SNN model.** The neurons were modelled using the quadratic integrate-and-fire model described previously.[44,54] In short, the single-neuron dynamics are described by the coupled differential equations:

$$\tau_v \dot{v} = a(v - v_{\text{ref}})(v - v_{\text{thr}}) - u + I_{\text{AMPA}} - I_{\text{GABA}} + \sigma\sqrt{2\tau_v}\xi,$$
$$\tau_u \dot{u} = b(v - v_{\text{ref}}) - u,$$

where $v$ and $u$ are variables describing the membrane potential and membrane recovery, respectively, with time constants $\tau_x$, and $\xi$ is a Gaussian noise term to model membrane fluctuations with an amplitude $\sigma$. Neurons interact through excitatory and inhibitory currents ($I_{\text{AMPA}}$ and $I_{\text{GABA}}$), which are described by a relaxation $\tau_x \dot{I_x} = -I_x$, which are increased at the postsynaptic neuron upon presynaptic firing by $I_{x,\text{post}} \to I_{x,\text{post}} + j_{x,\text{pre}} R_{\text{pre}}$, where $j_{x,\text{pre}}$ is a constant to describe the current strength, and $R$ is the presynaptic resource variable that decreases upon firing by $R \to \beta R$ and recovers as $\tau_D \dot{R} = 1 - R$. In addition, all neurons were spontaneously driven by an excitatory Poisson shot noise added to $I_{\text{AMPA}}$, which mimics both, spontaneous synaptic release and the optogenetic stimulation.

A modular network was constructed by considering four squares (200 μm × 200 μm) separated by 200 μm, locating 40 neurons randomly within each square, and simulating axon growth. Of these, 80% were excitatory neurons and 20% were inhibitory neurons. The number of intermodular connections $k$ was initially set, and the corresponding numbers of axons were forced to grow between each pair of modules. Binary adjacency matrices were then generated by forming synaptic connections when the axon of a presynaptic neuron intersected a circular region around a postsynaptic neuron within a radius of 150 ± 20 μm (mean ± SD). Full details of the model are provided in the Supplementary Section 2 (Supplementary Figs. 3–5).

**Mesoscopic model.** Each node $i$ in the mesoscopic model corresponds to a module, and its dynamics were modelled using a coupled rate model with resource depletion:[12]

$$\dot{\rho}_i(t) = -\frac{1}{\tau_\rho}\rho_i(t) + F[I_i(t)] + \sigma\xi_i(t),$$
$$\dot{R}_i(t) = -\frac{1}{\tau_{\text{d}}}\rho_i(t)R_i(t) + \frac{1}{\tau_{\text{c}}}\big(R_0 - R_i(t)\big),$$

where $\rho$ and $R$ are firing rate and synaptic resource variables, respectively, $R_0$ is the baseline resource level, and $\tau_x$ are time constants. $F(I_i)$ is a non-linear function mapping the total input to module $i$, $I_i$, to a rate change (Supplementary Section 3A). Modules were spontaneously driven by Gaussian noise $\xi$ with an amplitude $\sigma$, which was associated with internal biological variability.

Network models were constructed by coupling four modules together in a grid-like pattern (as in the 1-b and 3-b topologies), encoded by the adjacency matrix $A = [A_{ij}]$. Then, $I_i$ was the sum of external input $h$, activity propagation within the module, and activity propagation from connected neighbours:

$$I_i(t) = h + \rho_i(t)R_i(t) + w\sum_{j\neq i} A_{ij}g_{ij}(t)\rho_j(t)R_j(t),$$

where $w$ is the coupling strength, and $g_{ij}$ is the gating variable that describes whether modules $i$-$j$ are connected or disconnected. $h$ was varied to simulate perturbed conditions. The merged topology in the experiments corresponded to the behaviour of a single module unit. Further details of the model are provided in the Supplementary Section 3 (Supplementary Figs. 6–8).

**Data analysis.** For the analysis of collective activity events in the experimental data, the spike trains were first summed across all neurons and convolved with a normalized Gaussian kernel (SD = 200 ms), yielding a continuous time series that resembles a network-wide population rate. The start and end times of the events were then obtained by thresholding the population rate at 10% of the maximum observed for any recording. An event thus begins whenever the population rate exceeds the threshold and ends when the rate drops below the threshold. To account for fluctuations during an event, we also merged consecutive events if a start time was separated by less than 100 ms from a previous end time. Event size was then defined as the number of unique neurons that contributed to the event normalized by the total number of neurons in the network. Events in SNN models were defined analogously to experiments with the following parameters: SD of the Gaussian kernel = 20 ms and threshold = 2.5%. The adjustments were motivated by designing a kernel that scaled with the shortest observed inter-spike-interval and were necessary to account for the different sampling rates in simulations (5 ms) and experiments (50 ms).

The Pearson correlation coefficient $r_{ij}$ was used to quantify the synchronicity between a given pair of neurons $i$-$j$. For this analysis, the spike train was binned at 500 ms, and the number of spikes in each time-bin was counted for each neuron. From here, $r_{ij}$ was calculated by:

$$r_{ij} = \frac{\sum_t [x_i(t) - \bar{x}_i][x_j(t) - \bar{x}_j]}{\sqrt{\sum_t [x_i(t) - \bar{x}_i]^2}\sqrt{\sum_t [x_j(t) - \bar{x}_j]^2}}$$

where $x_i(t)$ is the time-binned spike train of neuron $i$ and $\bar{x}_i$ is the time-average of $x_i(t)$. Note that 'synchrony' and 'correlation' are used as synonyms throughout the manuscript.

The functional complexity $\chi$ was evaluated as:

$$\chi = 1 - \frac{m}{2(m-1)} \sum_{\mu=1}^{m} \left| p_\mu(r_{ij}) - \frac{1}{m} \right|,$$

where $p_\mu(r_{ij})$ is the probability distribution of $r_{ij}$ in bin $\mu$, $m$ = 20 is the number of bins for $r_{ij}$ used to estimate the distribution, and |.| denotes absolute value. The definition of the error bars is described in the captions of the corresponding figures.

**Data availability.** All data and codes that support the findings of this study will be made publicly available upon the acceptance of this manuscript.

## Acknowledgements

We thank K. Wakimura for assistance in setting up the instruments and running initial experiments on optogenetic stimulation and O. Vinogradov for insightful conversations. This research has been partly carried out at the Laboratory for Nanoelectronics and Spintronics, RIEC, Tohoku University, and the Fundamental Technology Center, RIEC, Tohoku University.

## Funding

HY, AHI, and SS acknowledge MEXT Grant-in-Aid for Transformative Research Areas (B) "Multicellular Neurobiocomputing" (21H05164), JSPS KAKENHI (18H03325, 20H02194, 20K20550, 22H03657), JST-PRESTO (JMPJPR18MB), JST-CREST (JPMJCR19K3), and

Tohoku University RIEC Cooperative Research Project Program for financial support. PS, VP, and JZ received support from the Max-Planck-Society. PS acknowledges funding by SMARTSTART, the joint training program in computational neuroscience by the VolkswagenStiftung and the Bernstein Network. VB was supported by a Sofja Kovalevskaja Award from the Alexander von Humboldt Foundation, endowed by the Federal Ministry of Education and Research. MAM acknowledges the Spanish Ministry and Agencia Estatal de investigación (AEI) through Project of I+D+i (PID2020-113681GB-I00) financed by MICIN/AEI/10.13039/501100011033 and FEDER "A way to make Europe", and the Consejería de Conocimiento, Investigación Universidad, Junta de Andalucía and European Regional Development Fund (P20-00173) for financial support. JZ received financial support from the Joachim Herz Stiftung. JS acknowledges Horizon 2020 Future and Emerging Technologies (964877-NEUChiP), Ministerio de Ciencia, Innovación y Universidades (PID2019-108842GB-C21), and Departament de Recerca i Universitats, Generalitat de Catalunya (2017-SGR-1061) for final support.

**Competing interests**

The authors declare no competing interests.

**Supplementary Information**

Supplementary Information file containing representative videos of experimental recordings and simulations, complementary analyses of experimental and numerical data as a well as a detailed description of both the microscopic and mesoscopic models are available at: https://www.mnbc.riec.tohoku.ac.jp/data/20220521_SI.pdf